\documentstyle[12pt,epsf]{article} 
\catcode`@=11
\newcount\@tempcntc
\def\@citex[#1]#2{\if@filesw\immediate\write\@auxout{\string\citation{#2}}\fi
  \@tempcnta\z@\@tempcntb\m@ne\def\@citea{}\@cite{\@for\@citeb:=#2\do
    {\@ifundefined
       {b@\@citeb}{\@citeo\@tempcntb\m@ne\@citea\def\@citea{,}{\bf ?}\@warning
       {Citation `\@citeb' on page \thepage \space undefined}}%
    {\setbox\z@\hbox{\global\@tempcntc0\csname b@\@citeb\endcsname\relax}%
     \ifnum\@tempcntc=\z@ \@citeo\@tempcntb\m@ne
       \@citea\def\@citea{,}\hbox{\csname b@\@citeb\endcsname}%
     \else
      \advance\@tempcntb\@ne
      \ifnum\@tempcntb=\@tempcntc
      \else\advance\@tempcntb\m@ne\@citeo
      \@tempcnta\@tempcntc\@tempcntb\@tempcntc\fi\fi}}\@citeo}{#1}}
\def\@citeo{\ifnum\@tempcnta>\@tempcntb\else\@citea\def\@citea{,}%
  \ifnum\@tempcnta=\@tempcntb\the\@tempcnta\else
   {\advance\@tempcnta\@ne\ifnum\@tempcnta=\@tempcntb \else \def\@citea{--}\fi
    \advance\@tempcnta\m@ne\the\@tempcnta\@citea\the\@tempcntb}\fi\fi}
\catcode`@=12

\begin{document}
\vskip -1.5cm
\begin{flushright}
ANL-HEP-PR-00-099 \\[-0.1cm]
CERN-TH/2000-267\\[-0.1cm]
EFI-2000-031 \\[-0.1cm]
FERMILAB-Pub-00/230-T\\[-0.1cm]
WUE-ITP-2000-027\\[-0.1cm] 
hep-ph/0009212\\[-0.1cm]
September 2000
\end{flushright}

\begin{center}
{\LARGE {\bf CP-Violating MSSM Higgs Bosons}}\\[0.3cm] 
{\LARGE {\bf in the Light of LEP~2}}\\[1.0cm]
{\large M. Carena$^{\,a}$, J. Ellis$^{\, b}$, A. Pilaftsis$^{\, c,b}$ 
and C.E.M. Wagner$^{\,d,e}$}\\[0.4cm]
$^a${\em Fermilab, P.O. Box 500, Batavia IL 60510, U.S.A.}\\[0.2cm]
$^b${\em Theory Division, CERN, CH-1211 Geneva 23, Switzerland}\\[0.2cm]
$^c${\em Institut f\"ur Theoretische Physik, Universit\"at W\"urzburg,\\
         Am Hubland, 97074 W\"urzburg, Germany}\\[0.2cm]
$^d${\em High Energy Physics Division, Argonne National Lab., Argonne
  IL 60439, U.S.A.}\\[0.2cm]
$^e${\em Enrico Fermi Institute, University of Chicago, 5640 Ellis Ave.,
Chicago
  IL 60637, U.S.A.}
\end{center}
\vskip0.7cm  \centerline{\bf ABSTRACT}  
In the MSSM, the CP parities of the neutral Higgs bosons may be mixed
by radiative effects induced by explicit CP violation in the third
generation of squarks.  To allow for this possibility, 
we argue that the charged Higgs-boson mass and $\tan\beta$ should be used to
parametrize the MSSM Higgs sector. We introduce a new benchmark
scenario of maximal CP violation appropriate for direct searches of
CP-violating MSSM Higgs bosons. We show that the bounds established by
LEP~2 on the MSSM Higgs sector may be substantially relaxed at low and
intermediate values of $\tan\beta$ in the presence of CP violation,
and comment on possible Higgs boson signatures 
at LEP~2 within this framework.

\newpage

The Minimal Supersymmetric extension of the Standard Model (MSSM)
constitutes the minimal viable scenario of low-energy supersymmetry
(SUSY), within which the problems of gauge hierarchy \cite{HPN} and
gauge-coupling unification \cite{CPW} can be successfully addressed.
Over the last 15 years, the MSSM has been the basis for many
theoretical \cite{Mh} and experimental Higgs-boson studies, serving as
a yardstick for models beyond the Standard Model (SM).  At the tree
level, the Higgs sector of the MSSM contains three neutral Higgs
bosons of definite CP parity: the CP-even Higgs bosons $h$ and $H$,
and the CP-odd Higgs scalar $A$ \cite{GHKD}. 
Radiative corrections to the spectrum and couplings of MSSM Higgs
bosons are important, leading in particular to an upper bound on the
lightest
neutral Higgs boson of about 130 GeV. Therefore, the presence of some
Higgs boson with a mass below about 130~GeV may be considered as a
`smoking gun' for the MSSM. At the time of writing, searches at LEP~2
quote a lower limit of 112.3~GeV on the mass of the SM Higgs boson,
excluding also a substantial part of the MSSM parameter space,
particularly at low $\tan\beta$~\cite{LEPC59}~\footnote{An excess of
  events consistent with a Higgs boson with SM-like couplings to gauge
  bosons and bottom quarks and a mass in the range 113--116 GeV has
  been reported at the 2.6 $\sigma$ level~\cite{LEPC59}. As a
  by--product of our analysis, we propose a novel interpretation for
  the origin of this possible excess.}.

It has been realized recently that the tree-level CP invariance of the
MSSM Higgs potential may be violated sizeably beyond the Born
approximation, by loop effects involving CP-violating interactions of
Higgs bosons to top and bottom squarks~\cite{APLB,PW,Demir,CDL,CEPW,KW}. 
In such a minimal SUSY scenario of explicit radiative CP violation, the
three neutral Higgs bosons, denoted by $H_1$, $H_2$ and $H_3$ in order of
increasing masses, have in general mixed CP parities.  It has been
found~\cite{PW,CDL,CEPW} that CP violation may modify drastically the
tree-level couplings of the Higgs particles to fermions and to gauge
bosons, thereby enabling~\cite{CEPW} even a relatively light Higgs boson
with $M_{H_1} \sim 60$~GeV to have escaped detection at LEP~2. 

We re-evaluate in this paper the physics potential of LEP~2 for
discovering Higgs bosons in the MSSM, in the presence of 
radiatively induced CP-violating effects in the Higgs sector.
We emphasize that, in the presence of CP-violating mixing
between the neutral Higgs bosons, it becomes necessary to parametrize
the MSSM Higgs sector in terms of the 
charged Higgs-boson mass $M_{H^+}$, since the commonly used CP-odd
Higgs boson mass is no longer associated with a physical Higgs mass
eigenstate.  The mass of the charged Higgs boson also controls the
strength of the CP-violating effects in the lightest Higgs 
sector~\cite{PW,CEPW}. For large values of $M_{H^+}$, the lightest 
neutral Higgs acquires SM-like properties  while the two heaviest
neutral Higgs bosons remain in general as states of mixed CP parity.

As a framework for the discussion, we introduce a new
benchmark scenario of maximal CP violation appropriate for analyzing the
direct searches for CP-violating Higgs bosons at LEP~2 and elsewhere.  We
then compare the MSSM Higgs discovery potential at LEP~2 for the
maximal CP-violating benchmark scenario and its CP-conserving counterpart.
We focus on low and intermediate values of $\tan\beta$, i.e., $\tan\beta
\stackrel{<}{{}_\sim} 7$, and on low values of $M_{H^+}$, $M_{H^+}
\stackrel{<}{{}_\sim} 170$ GeV, for which CP violation plays a particularly
important r\^ole. We find that the lightest neutral Higgs bosons may
be much lighter than the
quoted limit, and raise the possibility that the
apparent excess~\cite{LEPC59} may be due to the {\it second} neutral Higgs
boson.

Our numerical analysis is based on our earlier study in~\cite{CEPW},
in which we computed the one-loop renormalization-group (RG) improved
effective potential for the MSSM Higgs sector with explicit CP
violation.  Using RG methods, we calculated in~\cite{CEPW} the
charged and neutral Higgs-boson masses and couplings, including the
two-loop leading logarithms originating from QCD effects, as well as
those induced by the top- and bottom-quark Yukawa
couplings~\cite{CQW}. Also, we included the leading one-loop
logarithms associated with gaugino and higgsino quantum
effects~\cite{HH}.  Most importantly, we implemented the potentially
large two-loop non-logarithmic corrections induced by one-loop
threshold effects on the top- and bottom-quark Yukawa couplings, due
to the decoupling of the third-generation squarks~\cite{CHHHWW,zhang}.
The
numerical predictions presented here are obtained by the Fortran code
{\tt cph}~\cite{cph}, in which the aforementioned calculation of
Higgs-boson masses and couplings based on the RG-improved effective
potential has been implemented.

As  was  extensively  discussed   in~\cite{PW,CEPW},  there  are  two
important consequences of explicit radiative CP violation in the Higgs
sector. 
\begin{itemize}
  
\item The first is the generation of sizeable off-diagonal
  scalar-pseudoscalar contributions ${\cal M}^2_{SP}$~\cite{APLB,PW}
  to the general $3\times 3$ Higgs-boson mass matrix.  Each of the
individual CP-violating off-diagonal scalar-pseudoscalar mixing
entries ${\cal M}^2_{SP}$ in the neutral MSSM mass-squared matrix contains terms
scaling qualitatively as
\begin{equation}
  \label{MSP}
{\cal  M}^2_{SP}\ \sim\ \frac{m^4_t}{v^2}\, 
\frac{{\rm Im}\, (\mu A_t)}{32\pi^2\, M^2_{\rm SUSY}}\ 
\bigg( 1,\ \frac{|A_t|^2}{M^2_{\rm SUSY}}\,,\ 
\frac{|\mu|^2}{\tan\beta\,M^2_{\rm SUSY}}\,,\ 
\frac{2{\rm Re}\, (\mu A_t)}{M^2_{\rm SUSY}}\,\bigg)\, ,
\end{equation}
and could be of order $M^2_Z$.  In~(\ref{MSP}), $\mu$ is the
supersymmetric mixing parameter of the two Higgs superfields,
$M^2_{\rm SUSY}$ specifies the common soft SUSY-breaking scale defined
by the arithmetic average of the squared stop masses, and $A_t$ is the
soft SUSY-breaking trilinear coupling of the Higgs boson to top
squarks.

\item

The second important consequence of CP violation is the modification of
the top- and bottom-quark Yukawa couplings through CP-violating vertex
effects~\cite{PW,CEPW} involving gluinos and higgsinos, as well as top
and bottom squarks.  Although these effects enter the charged and neutral
Higgs-boson masses and couplings formally at the two-loop level, they can
still modify the numerical predictions or masses and couplings
in a significant way, and therefore have to be included in the analysis.

\end{itemize}

As we stressed above, the mass of the CP-odd Higgs scalar $A$ is no
longer an eigenvalue in the presence of CP violation, but rather just
one entry in the general $3\times 3$ neutral-Higgs mass matrix.
Therefore, in the general CP-violating case, it is no longer
appropriate to parametrize the MSSM Higgs sector in terms of $M_A$, as
is frequently done in the literature.  The mass of the charged Higgs
boson, $M_{H^+}$, instead, remains an observable physical parameter,
in terms of which the parametrization of the Higgs sector becomes
possible. As has been shown in~\cite{PW,CEPW}, all dominant
contributions to the neutral Higgs-boson mass matrix elements may be
expressed as a function of $M_{H^+}$, $\tan\beta$ and the soft
SUSY-breaking parameters associated with the third--generation
squarks, the $\mu$ parameter and the weak and strong gaugino masses.
Therefore, $M_{H^+}$ plays a very essential r\^ole in our analysis and
is a preferred replacement for $M_A$ as a physical MSSM Higgs
parameter.  Moreover, $M_{H^+}$ is also directly related to the mass
of the CP-odd Higgs boson $M_A$ in the CP-invariant limit of the
theory.  Therefore, $M_{H^+}$ is also an adequate physical input
parameter in the simplified case in which the three neutral Higgs
bosons carry definite CP parity.

It is obvious from (\ref{MSP}) that CP-violating effects on the
neutral Higgs-boson mass matrix become significant when the product
${\rm Im}\, (\mu A_t)/ M^2_{\rm SUSY}$ is large. Motivated by this
observation, we introduce the following new {\it CP-violating
  benchmark scenario (CPX)}:
\begin{eqnarray}
  \label{benchCP}
\widetilde{M}_Q \!&=&\! \widetilde{M}_t\ =\ 
\widetilde{M}_b\ =\ M_{\rm SUSY}\,,\qquad
\mu \ =\ 4 M_{\rm SUSY}\,,\nonumber\\
|A_t| \!&=&\!  |A_b|\ =\ 2M_{\rm SUSY}\,,\qquad 
{\rm arg}(A_t)\ =\ 90^\circ\,, \nonumber\\   
|m_{\tilde{g}}| \!&=&\! 1~{\rm TeV}\,,\qquad 
{\rm arg}(m_{\tilde{g}})\ =\ 90^\circ\, ,
\end{eqnarray}
where we follow the notation of~\cite{CEPW}.
Without loss  of generality, the  $\mu$ parameter is considered  to be
real. We note  that  the  CP-odd  angles  ${\rm  arg}(A_t)$  and  ${\rm
arg}(m_{\tilde{g}})$  are chosen to take their  maximal CP-violating
values. In the following, we also discuss variants of the CPX scenario
with other values of ${\rm arg}(A_t)$ and ${\rm arg}(m_{\tilde{g}})$,
keeping the other quantities fixed at the values in (\ref{benchCP}).

Large CP-odd phases are known to lead to
large contributions to the electron and neutron electric dipole
moments (EDMs)~\cite{EFN}. The main bulk of the large EDM effects may be
avoided by making the first two generation of squarks sufficiently
heavy, with masses of order 1~TeV and higher~\cite{PN}.  However, even
in this case, top and bottom squarks may give rise to observable
contributions to the electron and neutron EDMs through the three-gluon
operator~\cite{DDLPD}, through the effective coupling of the `CP-odd'
components of the 
Higgs boson to the gauge bosons~\cite{CKP}, and through two-loop
gaugino/higgsino-mediated EDM graphs~\cite{APino}, which may 
become large for large values of $\tan\beta$. However, exactly as
happens at the one-loop level~\cite{EFlores,IN}, these different EDM
contributions of the third generation can also have different signs
and add destructively to the electron and neutron EDMs. 
 
For our numerical comparisons, we also consider a related
CP-conserving benchmark scenario for which the stop mixing
parameters are chosen in order to maximize the lightest 
neutral Higgs boson mass value for large values of the charged 
Higgs mass (MAX)~\cite{CHWW}:
\begin{eqnarray}
  \label{bench}
\widetilde{M}_Q \!&=&\! \widetilde{M}_t\ =\ 
\widetilde{M}_b\ =\ M_{\rm SUSY}\,,\qquad
A_t\ =\  A_b\ =\ \sqrt{6}\, M_{\rm SUSY}\,,\nonumber\\   
m_{\tilde{g}} \!&=&\! 1~{\rm TeV}\,,\qquad
\mu\ =\ m_{\tilde{B}} \ =\ m_{\tilde{W}}\ =\ 200\ {\rm GeV}\, .
\end{eqnarray}
In  this CP-conserving  benchmark scenario,  we take  relatively small
values for the $\mu$ parameter and the gaugino masses $m_{\tilde{B}}$
and $m_{\tilde{W}}$, in  order to maximize the effect  of the one-loop
logarithmic   corrections   coming   from  chargino   and   neutralino
interactions~\cite{HH}.

We should stress that neither the MAX nor the CPX scenario are
generated in simple scenarios for SUSY breaking, such as
those based on minimal supergravity or gauge mediation. For example, in
supergravity models, the large values of
the $A_t$ parameter, compared to the third--generation squark masses, that
are needed in the MAX scenario, can
only be generated by large values of this parameter at the GUT scale,
an order of magnitude larger than the gaugino masses at that
scale.

The large values of $|\mu|$ chosen in the CPX scenario also
do not arise in minimal
supergravity or gauge-mediated models. Indeed, such large values of
$|\mu|$ can be consistent with electroweak symmetry breaking only if
the soft SUSY-breaking parameters associated with the Higgs
masses are negative and of the same order as $|\mu|$. This can 
easily be seen by ignoring
one-loop corrections, which are inessential for this specific discussion.
The soft SUSY-breaking mass parameters $m_1^2$ and $m^2_2$, which
are
associated with the scalar components of the Higgs-doublet superfields
$\widehat{H}_1$ and $\widehat{H}_2$, are then related to the value of
$\mu$ and the charged Higgs boson mass by the following relations:
\begin{eqnarray}
  \label{eq:mHc}
M_{H^+}^2  &=&   \frac{\tan^2\beta + 1}
{\tan^2\beta - 1}\, ( m_1^2 - m_2^2 )\ +\ M_W^2\ -\ M_Z^2\, ,\\
  \label{mu}
|\mu|^2 &=& \frac{m_1^2 - m_2^2 \tan^2\beta}{\tan^2\beta - 1}\
-\ \frac{M_Z^2}{2}\, .
\end{eqnarray}
We conclude from~(\ref{eq:mHc}) that, in order to get small values of
the charged Higgs mass, the values of $m_1^2$ and $m_2^2$ must be
close to each other.  In addition, it follows from (\ref{mu}) that
small values of $M_{H^+}$ are only compatible with large values of
$|\mu|$ if both the parameters $m_1^2$ and $m_2^2$ are large and
negative.

It is known that third-generation Yukawa-coupling effects may induce
negative values of the Higgs soft SUSY-breaking parameters resulting
in the breakdown of the electroweak symmetry.  
The large values of  the trilinear coupling at the high-energy
input scale which are required to generate $A_t \simeq 
{\cal{O}}(2 M_{\rm SUSY})$  at the weak scale, are helpful in driving 
$m_2^2$ to large negative values.
However, for small and moderate values of $\tan\beta$,
the large negative values of the soft SUSY-breaking Higgs mass parameter
$m_1^2$, 
necessary for the realization of the CPX scheme, can only be induced if
its value at the input scale is already negative and its
absolute value is larger than the squark masses. These non-standard
boundary conditions might be obtained, for instance, in models
inspired by superstring or $M$ theory, in which SUSY breaking is
induced by the vacuum expectation value of the auxiliary component of
moduli fields~\cite{Mtheory}.

The aim of the MAX and CPX benchmark scenarios, however, is to study
the phenomenological consequences for Higgs searches for the most
challenging values of the MSSM parameters. This allows one to study
the capability of the present and near future colliders to explore the
Higgs boson properties in the most generic framework.

\begin{figure}
   \leavevmode
 \begin{center}
 \vspace{-3.cm}
   \epsfxsize=18.cm
    \epsffile[0 0 539 652]{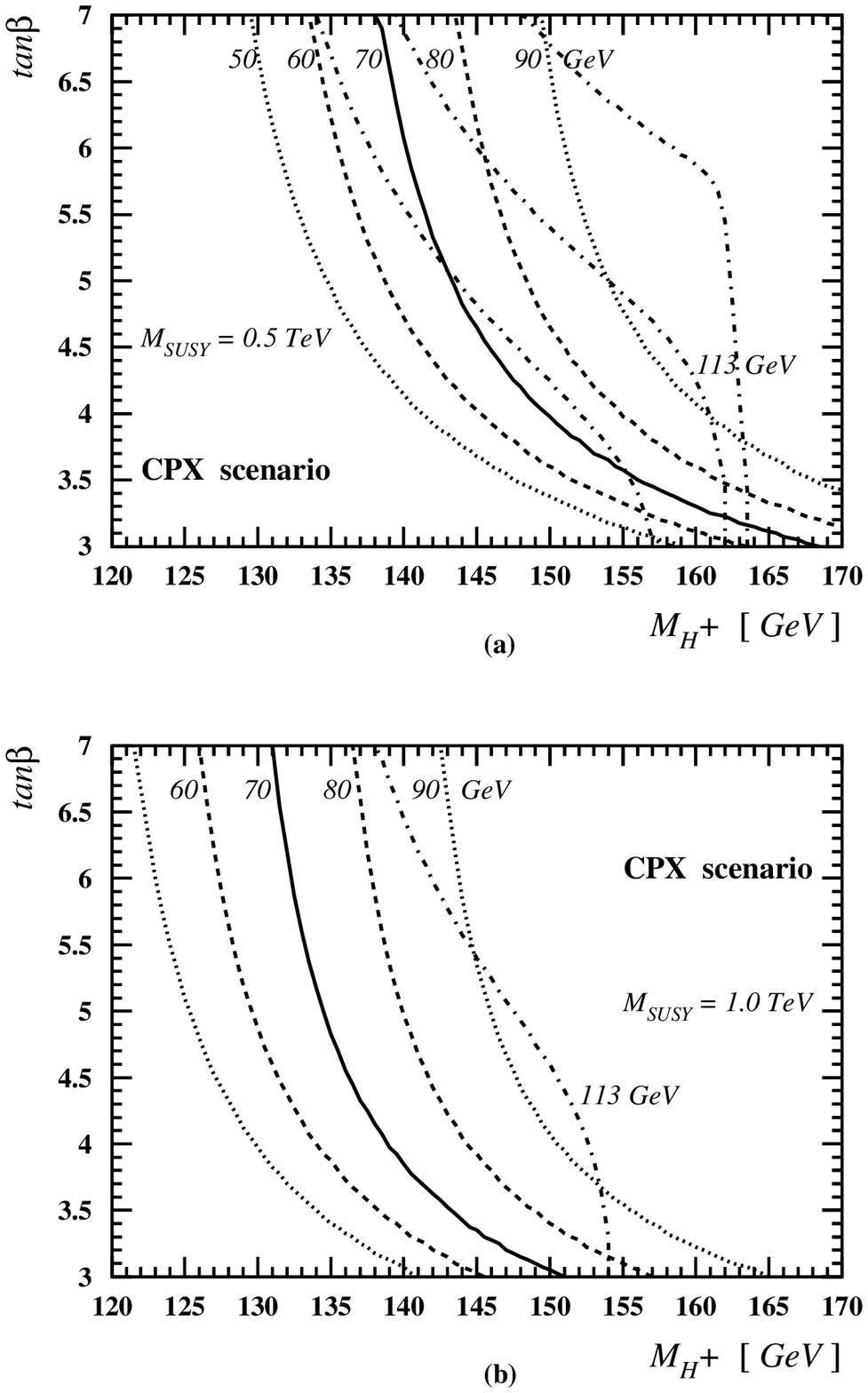}
 \end{center}
 \vspace{-1.5cm} 
\caption{\it Contours  of the lightest neutral Higgs
  boson mass $M_{H_1}$ in the $(M_{H^+},\tan\beta )$ plane, for 
  the CP-violating benchmark scenario~(\ref{benchCP}) CPX and
values of $M_{\rm SUSY} = 0.5$ and 1 TeV. Contours of $M_{H_2}=
113$~GeV are also indicated by dash-dotted lines. The three
such lines in the upper panel correspond (from right to left) to $m_t =
170, 175, 180$~GeV: all the other contours in this and subsequent
figures are drawn for our default choice $m_t = 175$~GeV.}
\label{fig:masscontours}
\end{figure}

We display in Fig.~\ref{fig:masscontours} contours of the masses of
the two lightest neutral Higgs bosons $H_{1,2}$ in the CPX scenario,
for the two values $M_{\rm SUSY} = 0.5$, 1 TeV.  As we shall see,
neutral Higgs bosons as light as about 50~GeV are allowed within the
CPX scenario, whereas such a light Higgs boson is not allowed in the
MAX scenario.  In fact, there are significant regions of parameter
space in the $(M_{H^+}, \tan \beta)$ plane where the lightest neutral
Higgs boson $H_1$ contains a large admixture of the CP-odd state $A$.
In this case, there are important consequences for the $H_1$
couplings, as we now discuss.

At LEP~2, the main  production mechanism for  the neutral  Higgs bosons
$H_i$ and $i=1,2,3$ is the  Higgs-strahlung process:  $e^+ e^-\to
Z^*\to Z  H_i$ \cite{ZH}. If  the neutral Higgs bosons  are relatively
light, they may  also be produced in pairs  through the reaction: $e^+
e^-\to Z^*\to H_i H_j$, with  $i\neq j$ and $i,j = 1,2,3$.  Therefore,
the  interaction Lagrangian  of  interest to  us,  which describes  the
effective $H_iZZ$ and $H_iH_jZ$ couplings, is given by
\begin{eqnarray}
  \label{HVV}
{\cal L}_{\rm int} &=& \frac{g_w}{2\cos\theta_w}\, \bigg[\,M_Z\, 
\sum\limits_{i=1}^3\, g_{H_iZZ}\, H_i Z_\mu Z^\mu\: +\: 
\sum\limits_{j>i=1}^3\, g_{H_iH_jZ}\, ( H_i\, \!\! 
\stackrel{\leftrightarrow}{\vspace{2pt}\partial}_{\!\mu} H_j
)\,Z^\mu\, \bigg]\, ,
\end{eqnarray}
where              $\cos\theta_w              \equiv M_W/M_Z$,
$\stackrel{\leftrightarrow}{\vspace{2pt}  \partial}_{\!  \mu}\ \equiv\ 
\stackrel{\rightarrow}{\vspace{2pt}      \partial}_{\!      \mu}     -
\stackrel{\leftarrow}{\vspace{2pt} \partial}_{\! \mu}$, and
\begin{eqnarray}
  \label{gHZZ}
g_{H_iZZ} &=& \cos\beta\, O_{1i}\: +\: \sin\beta\, O_{2i}\,,\nonumber\\  
 \label{gHHZ}
g_{H_i H_j Z} &=& O_{3i}\, \Big( 
\cos\beta\, O_{2j}\, -\, \sin\beta\, O_{1j}  \Big)\ -\
 O_{3j}\, \Big(  \cos\beta\, O_{2i}\,  -\,  \sin\beta\, O_{1i} \Big)\, .
\end{eqnarray}
Here, $\tan\beta  = v_2/v_1$  is the ratio  of the  vacuum expectation
values  of the two  Higgs doublets,  and $O$  is an  orthogonal matrix
relating the weak to the mass eigenstates of the neutral Higgs bosons.
The effective couplings  $H_i ZZ$ and $H_i H_j Z$  are related to each
other through
\begin{equation}
  \label{Orel}
g_{H_k ZZ}\ =\ \varepsilon_{ijk}\, g_{H_i H_j Z}\, . 
\end{equation}
Unitarity leads to the coupling sum rule~\cite{Alex}
\begin{equation}
\sum\limits_{i=1}^3\, g_{H_i ZZ}^2\ =\ 1\, ,
\end{equation}
which reduces the number of the independent $H_iZZ$ and $H_iH_jZ$
couplings. Finally, there is another important sum rule involving the
neutral Higgs-boson masses and their respective couplings to the $Z$
bosons:
\begin{equation}
  \label{coupl}
\sum\limits_{i=1}^3\, g_{H_i ZZ}^2\, M^2_{H_i}\ =\
M^{2,\,{\rm max}}_{H_1}\, ,
\end{equation}
where $M^{2,\,{\rm max}}_{H_1}$ is the $H_1$-boson mass in the
decoupling limit, in which $M_{H^+}\gg 2M_Z$ and all other parameters,
apart from the charged-Higgs mass, are assumed to be the same in the
computation of both sides of (\ref{coupl}).  The mass-coupling sum
rule (\ref{coupl}) is very analogous to the one found in \cite{CMW}
for the CP-conserving case.  In the decoupling limit, we have
$g^2_{H_1 ZZ} \to 1$, so the lower limit on $M_{H_1}$ reaches its
maximum value.

\begin{figure}
   \leavevmode
 \begin{center}
 \vspace{-2.cm}
   \epsfxsize=18.cm
    \epsffile[0 0 539 652]{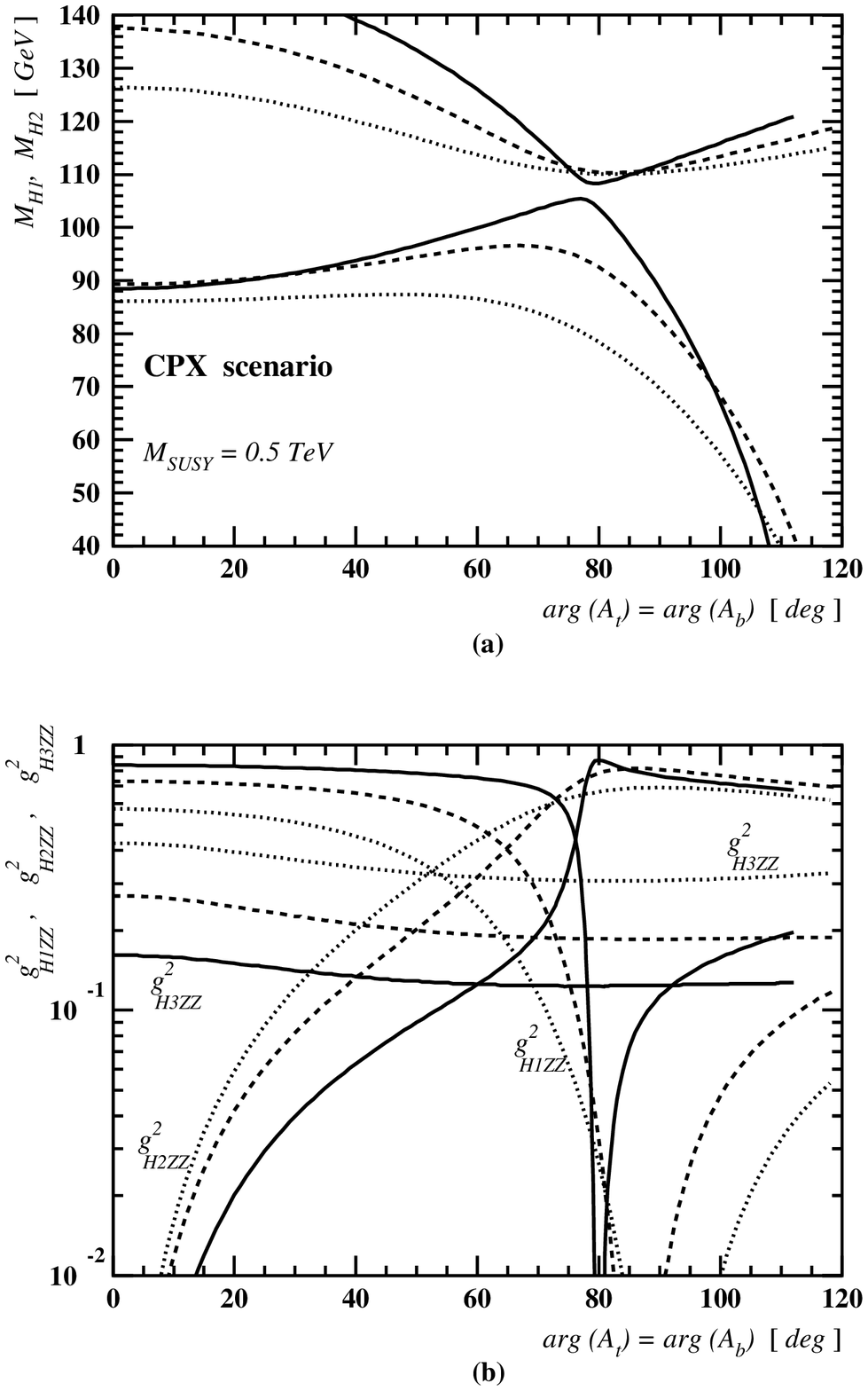}
 \end{center}
 \vspace{-1.cm} 
\caption{\it Predicted values  of (a) $M_{H_1}$ and $M_{H_2}$ and
(b) $g^2_{H_iZZ}$ as functions of ${\rm arg}\, (A_t)$, in the
CPX scenario for $M_{\rm SUSY} = 1$ TeV and 
for the
following choices of $(M_{H^+}, \tan\beta )$: (160~GeV, 4) (solid
lines), (150~GeV, 5) (dashed lines) and (140~GeV, 6) (dotted lines).}
\label{fig:ZZH}
\end{figure}

We display in Fig.~\ref{fig:ZZH} the masses $M_{H_i}$ of the two
lightest neutral Higgs bosons, as well as their corresponding
couplings $g_{H_i ZZ}$ to the $Z$ boson for the soft supersymmetry
breaking parameters defined in the CPX scenario for several choices of
$(M_{H^+}, \tan\beta)$, varying the CP-violating phases
$arg(A_{t,b})$.  The level-crossing phenomenon discussed
in~\cite{CEPW} is clearly visible in the upper panel, and is
associated with a change in the strength of the couplings of the
lightest and next-to-lightest Higgs bosons to the $Z$ gauge boson, as
seen in the lower panel. As is apparent in the lower panel, the
$g_{H_1 ZZ}$ coupling is strongly suppressed for values of $M_{H^+}$
and $arg(A_{t,b})$ near those where the level crossing takes place.

\begin{figure}
   \leavevmode
 \begin{center}
 \vspace{-2.cm}
   \epsfxsize=18.cm
    \epsffile[0 0 539 652]{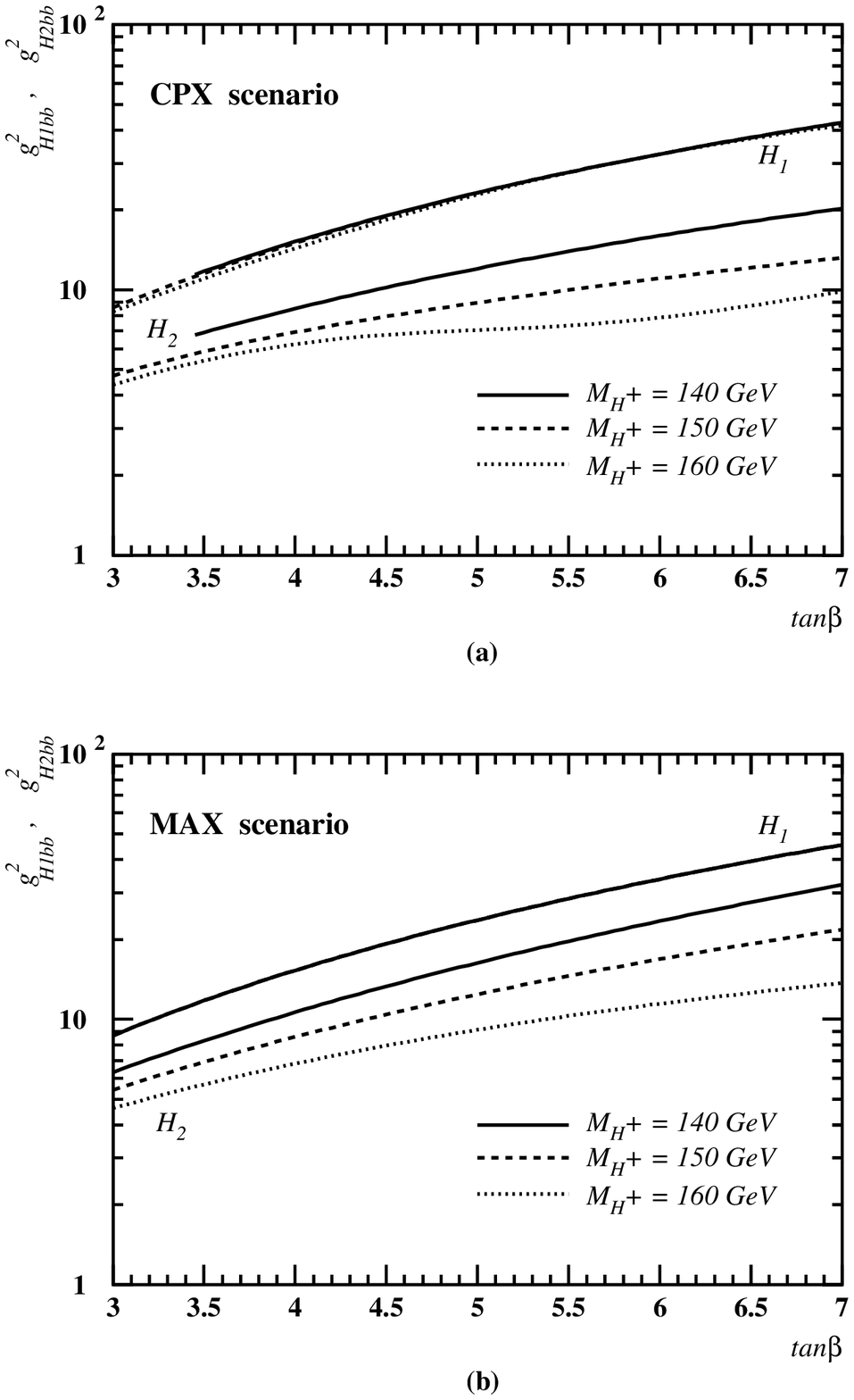}
 \end{center}
 \vspace{-1.cm} 
\caption{\it Numerical estimates of the squared $H_1b\bar{b}$ and
  $H_2b\bar{b}$ couplings, normalized to their SM values,
as functions of $\tan\beta$, in the CP-violating benchmark scenario
  (\ref{benchCP}) CPX (top panel) and the maximal stop-mixing
CP-conserving scenario (\ref{bench}) MAX (bottom panel).}
\label{fig:bbbarH}
\end{figure}

On the other hand, the branching ratios for $H_{1,2} \rightarrow {\bar
  b} b$ decay are not in general greatly modified in comparison with
the CP-conserving MAX scenario, as can be seen in
Fig.~\ref{fig:bbbarH}.  This property is expected to be valid for low
values of $\tan\beta$, such as the ones considered here.  For larger
values of $\tan\beta$, scalar-pseudoscalar mixing effects induced by
stop and sbottom quantum effects become relatively less
significant\footnote{CP-violating chargino effects were recently
  computed in \cite{INrecent}, and found to be of relevance only for
  large values of $\tan\beta \stackrel{>}{{}_\sim} 30$. They are
  comparable with the small squark effects of the third generation in
  this large-$\tan\beta$ regime.}, and CP-violating vertex effects on
Higgs-boson decays become more important \cite{CEPW,CL}. Consistent
with Fig.~\ref{fig:bbbarH}, in what follows we assume that the
branching ratios for the decays of the $H_1$ and $H_2$ into ${\bar
  b}b$ are $\simeq 1$, and pursue the implications of the differences
in production cross sections for $e^+ e^- \rightarrow Z + H_{1,2}$ and
$e^+ e^- \rightarrow H_1 H_2$ between the MAX and CPX benchmarks.

\begin{figure}
   \leavevmode
 \begin{center}
 \vspace{-3.cm}
   \epsfxsize=18.cm
    \epsffile[0 0 539 652]{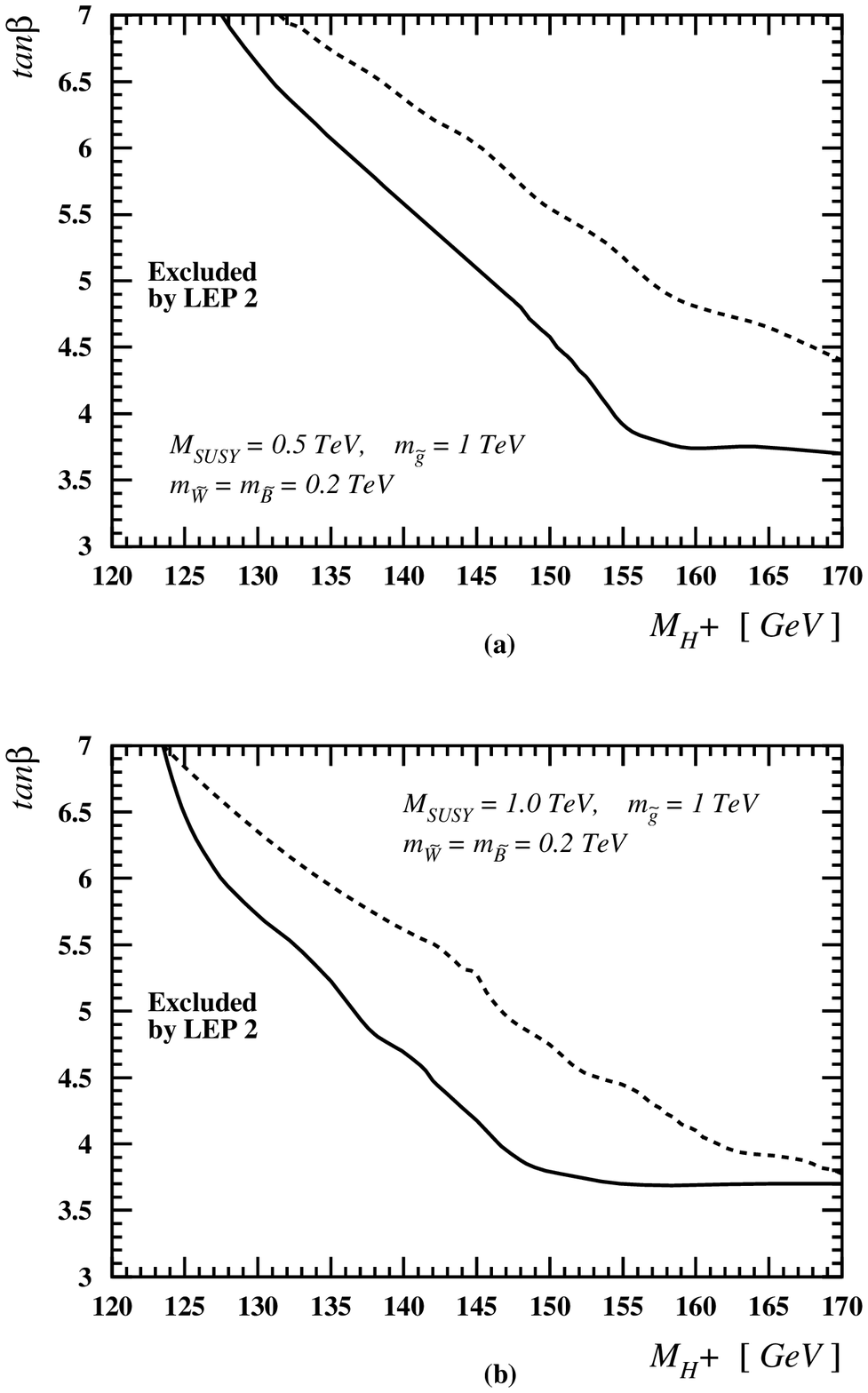}
 \end{center}
 \vspace{-1.5cm} 
\caption{\it Approximate 95 \% C.L. exclusion plots in the
$(M_{H^+},\tan\beta)$ plane, 
for the CP-violating benchmark scenario (\ref{benchCP}) CPX (solid
lines) and the maximal stop-mixing CP-conserving scenario (\ref{bench})
MAX (dashed lines), for the two indicated sets of soft
SUSY-breaking parameters. }
\label{fig:lep2}
\end{figure}

In Fig.~\ref{fig:lep2} we compare the 95\% confidence-level exclusion
limits~\cite{exp,Read} on the neutral Higgs bosons in the $(M_{H^+},
\tan\beta)$ plane for the two scenarios CPX and MAX, for the choices
$M_{\rm SUSY} = 0.5$ and 1 TeV in the upper and lower panels,
respectively. The solid lines in Fig.~\ref{fig:lep2} refer to the
CP-violating CPX benchmark scenario (\ref{benchCP}), whilst the dashed
ones are for the CP-conserving MAX benchmark scenario (\ref{bench}).
In order to obtain the limits shown, we have rescaled the quoted SM
Higgs mass limits~\cite{Read} to take account of the fact that no
SM-like Higgs boson has yet been observed with a mass up to about
112.3~GeV at the 95\% confidence level (CL)~\cite{Read}~\footnote{ The
  irregularities of the lines originate from our approximate reading
  of the experimental limits~\cite{Read}. Our intention here is to
  establish the differences between the CPX and MAX benchmark
  scenarios. We leave more complete studies to the LEP Collaborations
  and the LEP Higgs Working Group.}.  The areas lying to the left of
the lines are excluded. We concentrate on small values of the
charged-Higgs-boson mass~\footnote{However, we respect the direct
  experimental limits for $M_{H^+}$ established at LEP~2 and the Tevatron
  \cite{Tevatron}.}, for which the effect of CP violation is
maximized.

As we mentioned above, for large values of the charged-Higgs-boson
mass, CP-violation effects decouple from the lightest neutral Higgs
boson, whose couplings to fermions and gauge bosons resemble those of
the SM Higgs boson.  Therefore, for large values of the
charged-Higgs-boson mass and for fixed values of the third--generation
squark--mass parameters, the MAX scenario leads to the most
conservative limits on $\tan\beta$.  For smaller values of the charged
Higgs mass, instead, CP-violating effects can considerably weaken the
quoted LEP Higgs bound. Indeed, we see in Fig.~\ref{fig:lep2} that,
for any given value of $\tan\beta$ and $M_{\rm SUSY}$, lower values of
$M_{H^+}$ are allowed in the CPX scenario.  Comparing with
Fig.~\ref{fig:masscontours}, we see that, due the effects of CP
violation the lightest neutral Higgs boson $H_1$ could be as light as
about 50~GeV and remain undetected at LEP.

We also observe in Figs~\ref{fig:masscontours} and \ref{fig:ZZH}
another interesting feature, namely that $M_{H_2}$ lies between 110
and 120~GeV over much of the parameter region where $M_{H_1} <
100$~GeV~\footnote{At present, the theoretical uncertainties in the
  mass calculation are of about 3~GeV ~\cite{CHHHWW,zhang} and,
  moreover, there are uncertainties due to the value of $m_t$, which
  may lie in the range 170 to 180 GeV. The top panel of
  Fig~\ref{fig:masscontours} illustrates the corresponding uncertainty
  in the $(M_{H^+}, \tan\beta)$ plane.}.  In this region, the lightest
neutral Higgs boson may remain unobserved due to a strong suppression
of its coupling to the $Z$ gauge boson.  The second-lightest neutral
Higgs-boson, instead, has couplings to gauge bosons and to bottom
quarks that are similar to those in the SM, and its mass may be
consistent with the apparent excess of events reported recently at
LEP~2~\cite{LEPC59}. Therefore, it is conceivable that LEP has
evidence for the {\it second-lightest} Higgs boson, not the
lightest~\footnote{A second-lightest Higgs boson with SM-like
  couplings to gauge bosons and fermions can also appear in the
  absence of CP violation~\cite{CMW}. However, in this case, it tends
  to occur for larger values of $\tan\beta$, for which the
  second-lightest Higgs boson is typically heavier, with mass larger
  than 115 GeV.}, even though this may not be the most obvious
interpretation.

On a more speculative note, we observe that the combination of the
data of the four LEP experiments has not only led to an apparent
excess around 115 GeV, but may also not be able to exclude an
additional excess of events around 95 GeV at a level corresponding to
a Higgs cross section about a tenth of the SM cross
section~\cite{Junk}.  As is shown in Fig.~\ref{fig:ZZH}, both
`excesses' could be reproduced simultaneously in the CP-violating
scenario, for $M_{H^+} \sim 160$~GeV, $\tan\beta \sim 4$ and $arg(A_t)
\sim$ 85 degrees.

In conclusion: we have demonstrated how radiative corrections in the
MSSM with explicit CP violation can enlarge the parameter space
allowed for the MSSM Higgs sector at low values of $\tan\beta$ and the
charged Higgs-boson mass $M_{H^+}$, i.e., for $\tan\beta
\stackrel{<}{{}_\sim} 7$ and $M_{H^+} \stackrel{<}{{}_\sim} 170$~GeV.
In developing this analysis, we have introduced a new CP-violating
benchmark scenario for Higgs mixing, which should be useful for
testing the CP-violating scenario at LEP2 and other colliders.
We have also emphasized that the
optimal parametrization of MSSM Higgs bosons is in terms of
$M_{H^\pm}$ and $\tan\beta$. Finally, as a by-product, we have given a
novel
interpretation of the possible excess of events recently reported
at LEP~2~\cite{LEPC59}.
If this excess were to be confirmed by further LEP running,
data from the Tevatron collider and from the LHC would
be necessary in order to discriminate this possibility from more
standard explanations.

\subsection*{Acknowledgements}
AP thanks the Theory Group of Argonne National Laboratory for its warm
hospitality while part of the work was being done. M.C. and C.W.
would like to thank T. Junk for useful discussions.  Work supported
in part by the US Department of Energy, High Energy Physics Division,
under Contract W-31-109-Eng-38.

%
%

\end{document}